\documentstyle[psfig,conf_iap]{article}
\newcommand{\beq}{\begin{equation}}
\newcommand{\eeq}{\end{equation}}
\newcommand{\beqa}{\begin{eqnarray}}
\newcommand{\eeqa}{\end{eqnarray}}

\def\d{\delta}

\def\ds{\delta_s}

\font\BF=cmmib10

\def\k{{\hbox{\BF k}}}
\def\x{{\hbox{\BF x}}}

\def\v{{\hbox{\BF v}}}
\def\u{{\hbox{$u_z$}}}

\def\fun#1#2{\lower3.6pt\vbox{\baselineskip0pt\lineskip.9pt
        \ialign{$\mathsurround=0pt#1\hfill##\hfil$\crcr#2\crcr\sim\crcr}}}
\begin{document}
%
\heading{Redshift Distortions: Perturbative and N-body Results} 
\par\medskip\noindent
\author{Rom\'{a}n Scoccimarro}
\address{%
CITA, McLennan Physical Labs, 60 St George, Toronto, ON M5S~3H8, Canada} 

\begin{abstract}
I discuss the evolution of the redshift-space bispectrum via
perturbation theory (PT) and large high-resolution numerical
simulations. At large scales, we give the multipole expansion of the
bispectrum in PT, which provides a natural way to break the degeneracy
between bias and $\Omega$ present in measurements of the power
spectrum distortions. At intermediate scales, we propose a simple
phenomenological model to take into account non-linear effects. N-body
results show that at small scales the perturbative shape of the
bispectrum monopole in redshift-space is preserved, breaking the
hierarchical form valid in the absence of distortions.
\end{abstract}
%
%

%
\section{Introduction and PT Results}
%
The bispectrum, the three-point function of density fluctuations in
Fourier space, is the lowest order statistic that carries information
about the spatial coherence of large-scale structures. Non-linear
perturbation theory predicts a characteristic dependence of the
bispectrum on the shape of the triangle, which provides a signature of
gravitational instability that can be used to probe the gaussianity of
initial conditions and the bias of the galaxy distribution
\cite{Fry94}. However, in order to use it in redshift surveys,
redshift distortions due to peculiar velocities must be taken into
account. In this talk, I discuss the evolution of the redshift-space
bispectrum in PT and N-body simulations \cite{SCF98}.

In the plane-parallel approximation, where the observed position ${\bf
s}$ of a galaxy is given by ${\bf s}=\x - f \ \u(\x) {\hat z}$, with
${\hat z}$ a fixed direction, $f(\Omega) \approx \Omega^{0.6}$, and
${\bf u}(\x) \equiv - \v(\x)/({\cal H} f)$, with $\v(\x)$ the peculiar
velocity field; the density contrast in redshift-space reads
\cite{SCF98}
\begin{equation} 
\label{d_s}
\ds(\k) = \int \frac{d^3x}{(2\pi)^3}\ {\rm e}^{-i \k\cdot\x}\ {\rm e}^{i
f k_z \u(\x)}\ \Big[ \d(\x) + f \nabla_z \u(\x) \Big].
\end{equation}
This equation describes the full non-linear density field in
redshift-space in the plane-parallel approximation, and is the
starting point for the perturbative approach in redshift-space. The
term in square brackets describes the ``squashing effect'', i.e. the
increase in clustering amplitudes due to infall, and gives the
standard Kaiser formula in linear PT \cite{Kaiser87}. The exponential
factor encodes the ``fingers of god'' (FOG) effect, which erases power
due to velocity dispersion along the line of sight.  From
Eq.~(\ref{d_s}), it is straightforward to obtain the density field to
any order in PT; in particular, the hierarchical amplitude $Q_s$
follows from second-order PT \cite{HBCJ95}

\begin{equation}
Q_s(\k_1,\k_2,\k_3) \equiv \frac{B_s(\k_1,\k_2,\k_3)}{[P_g^0(k_1) \
P_g^0(k_2) + P_g^0(k_2) \ P_g^0(k_3) + P_g^0(k_3) \ P_g^0(k_1)]}
\end{equation}
where the $B_s$ denotes the bispectrum and $P_g^0(k)\equiv b^2 (1+
\frac{2}{3}\beta+\frac{1}{5}\beta^2) P(k)$ is the power spectrum
monopole of galaxies in linear PT assuming deterministic biasing, and
$\beta\equiv f/b$. Decomposing into multipoles with respect to $\mu$,
where $\mu \equiv \hat{\k}_1 \cdot \hat{z}$, we obtain the tree-level
monopole of the equilateral hierarchical amplitude \cite{SCF98}:
\begin{equation}
\label{Qs0}
Q_{s\ {\rm eq}}^{(\ell=0)}= 5\ A(\gamma,\beta,b)\ / 
[ 98\,b\, ( 15 +10\,\beta + 3\,\beta^2 )^2], 
\end{equation}

\noindent where $A\equiv 2520+4410\gamma + 3360\beta +
2940\gamma\beta + 1260\beta^2 + 441\gamma\beta^2 + 72\beta^3 -
63b\beta^3 - 14b\beta^4$, $\gamma \equiv b_2/b$, and $b_2$ is the
non-linear bias.  For no biasing this yields $Q_{s\ {\rm
eq}}^{(0)}=0.464$ for $\Omega=1$. The quadrupole to monopole ratio of
$B_{\rm eq}$ is
\begin{equation}
\label{RBeq}
R_{\rm B} =
[5/(22 A)] ( 7392\beta + 6468\gamma\beta + 
3960\beta^2 + 1386\gamma\beta^2 + 264\beta^3 - 
231b\beta^3 - 56b\beta^4 ),  
\end{equation}
which for no biasing gives $R_{\rm B}=11329/31394=0.36$ for
$f=1$. Note that Eqs.~(\ref{Qs0}-\ref{RBeq}) give a constraint on
$(b,\beta,\gamma)$ that is independent of the power spectrum. These,
together with $R_P$, the power spectrum quadrupole to monopole ratio
statistic \cite{Hamilton92,CFW94}, can be inverted to obtain $\Omega$,
$b$ and $b_2$ at large scales, independent of the initial conditions.

%
\section{Non-Linear Redshift Distortions and N-Body Results}
%

In order to describe the non-linear behavior of the redshift-space
bispectrum, we introduce a phenomenological model to take into account
the effects of velocity dispersion . For the power spectrum, we take
\cite{PVGH94} 
\begin{equation}
\label{Ppheno}
P_s(\k)= b^2\ P(k)\ (1+\beta \mu^2)^2\ /[1+(k\mu\sigma_v)^2/2]^2,
\end{equation}
where $\sigma_v$ is a free parameter that characterizes the velocity
dispersion along the line of sight.  It is straightforward to obtain
the multipole moments of $P_s(\k)$ in this simple model. Here we use
the quadrupole to monopole ratio statistic, $R_P$, to fit $\sigma_v$,
and then propose a similar ansatz to take into account non-linear
distortions of the bispectrum:
\begin{equation}
\label{Bpheno}
B_s(\k_1,\k_2,\k_3)= B_s^{\rm PT}(\k_1,\k_2,\k_3)\ /\big[1+\alpha^2\
[(k_1\mu_1)^2 + (k_2\mu_2)^2 + (k_3\mu_3)^2]^2 \sigma_v^2/2\big]^2,
\end{equation} 
where $B_s^{\rm PT}(\k_1,\k_2,\k_3)$ is the tree-level redshift-space
bispectrum. Note that we introduced a
constant $\alpha$ which reflects the configuration dependence of the
triplet velocity dispersion.  At $z=0$, we take $\alpha \simeq 2$ for
equilateral configurations, and $\alpha \simeq 3$ for $k_1/k_2=2$
configurations, independent of cosmology. 

Figure~1 shows the results from this modeling and how it compares with
N-body simulations, corresponding to $256^3$ particles in a
240$h^{-1}$ Mpc box run by the Virgo Consortium.  For the SCDM model,
from $R_{\rm P}$ we obtain $\sigma_v=6$ (in units of H$_0$=100 $h$
km/s/Mpc) for $z=0$ and $\sigma_v=2$ for z=1 (with $\alpha=1$). Using
this we can predict $R_{\rm B}$ by taking multipoles in
Eq.~(\ref{Bpheno}), the resulting $R_{\rm B}$ is shown as solid lines
in Fig.~2, which despicts excellent agreement with the N-body
simulation results. Similarly, for the $\Lambda$CDM model we obtain
from fitting $R_{\rm P}$, $\sigma_v=5.5 $ for $z=0$ and $\sigma_v=4$
for $z=1$ (with $\alpha=1.75$). The results for $R_{\rm B}$ are also
in very good agreement with numerical simulations.

The bottom panels in Fig.~1 show a comparison of simulations to the
predictions of PT and the model of Eq.~(\ref{Bpheno}) in $k_2/k_1=2$
configurations, for three different scales. We see in these panels
that even though the tree-level PT prediction in real space (dotted)
works reasonably well, the redshift-space counterpart (dashed) does
not. On the other hand, the model in Eq.~(\ref{Bpheno}) (solid)
describes the N-body results very well. Even at the largest scale we
probed, corresponding to a wavelength $\lambda \simeq 50$ Mpc/$h$, the
tree-level PT prediction in redshift-space would predict an effective
bias $b=1.4$, a quite significant discrepancy.  This situation is
similar to what happens with the power spectrum, redshift-space
statistics are more affected by non-linearities than their real-space
counterparts.  In this respect, it is interesting to note that even in
the linear dynamics, the exponential factor in Eq.~(\ref{d_s}) can
lead to a FOG effect at large scales that eventually makes $R_{\rm P}$
and $R_{\rm B}$ to become negative \cite{SCF98}. Therefore, the long
range of the FOG effect seen in numerical simulations, should not be
exclusively attributed to virialized clusters. These results strongly
suggest the possibility of extending the leading-order PT results for
the power spectrum and bispectrum to smaller scales by treating the
redshift-space mapping in Eq.~(\ref{d_s}) exactly and approximating
the dynamics using PT \cite{SCF98}.  The bottom right panel in Fig.~1
nicely illustrates the effect of cluster velocity dispersion on
redshift-space correlations in the non-linear regime, whereas $Q$ is
very close to hierarchical, $Q_s$ has a strong configuration
dependence. Eq.~(\ref{Bpheno}) (solid) does an excellent job in
predicting $Q_s$, even at this considerable stage of non-linearity.

\acknowledgements{This material is based on a paper in collaboration
with H. Couchman and J. Frieman \cite{SCF98}. The N-body 
simulations were carried out by the Virgo
Supercomputing Consortium
(http://star-www.dur.ac.uk/~frazerp/virgo/virgo.html) using computers
based at the Max Plank Institut fur Astrophysik, Garching and the
Edinburgh Parallel Computing Centre.  }

\begin{figure}
\centerline{\vbox{
\psfig{figure=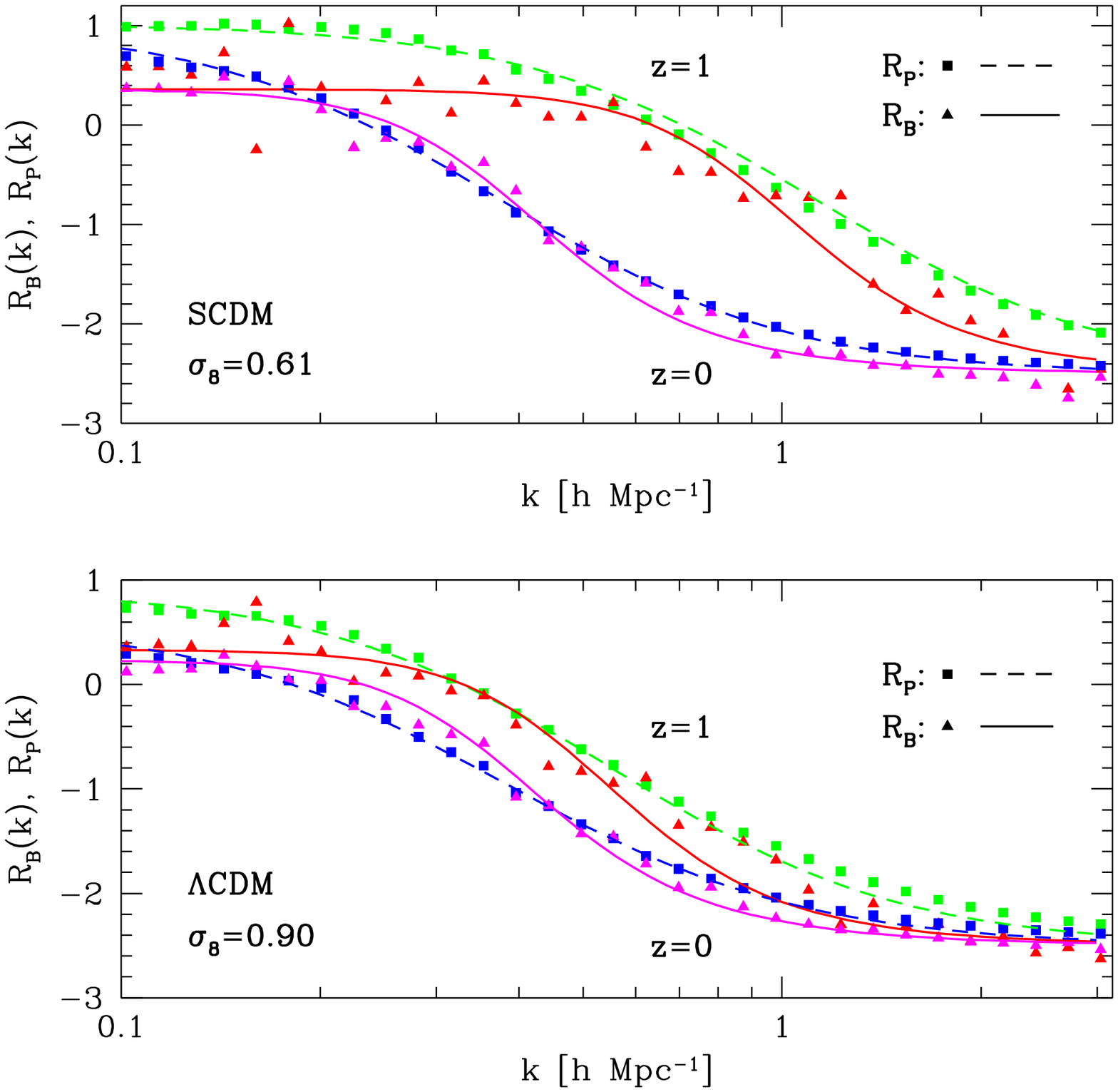,height=8.cm}
\psfig{figure=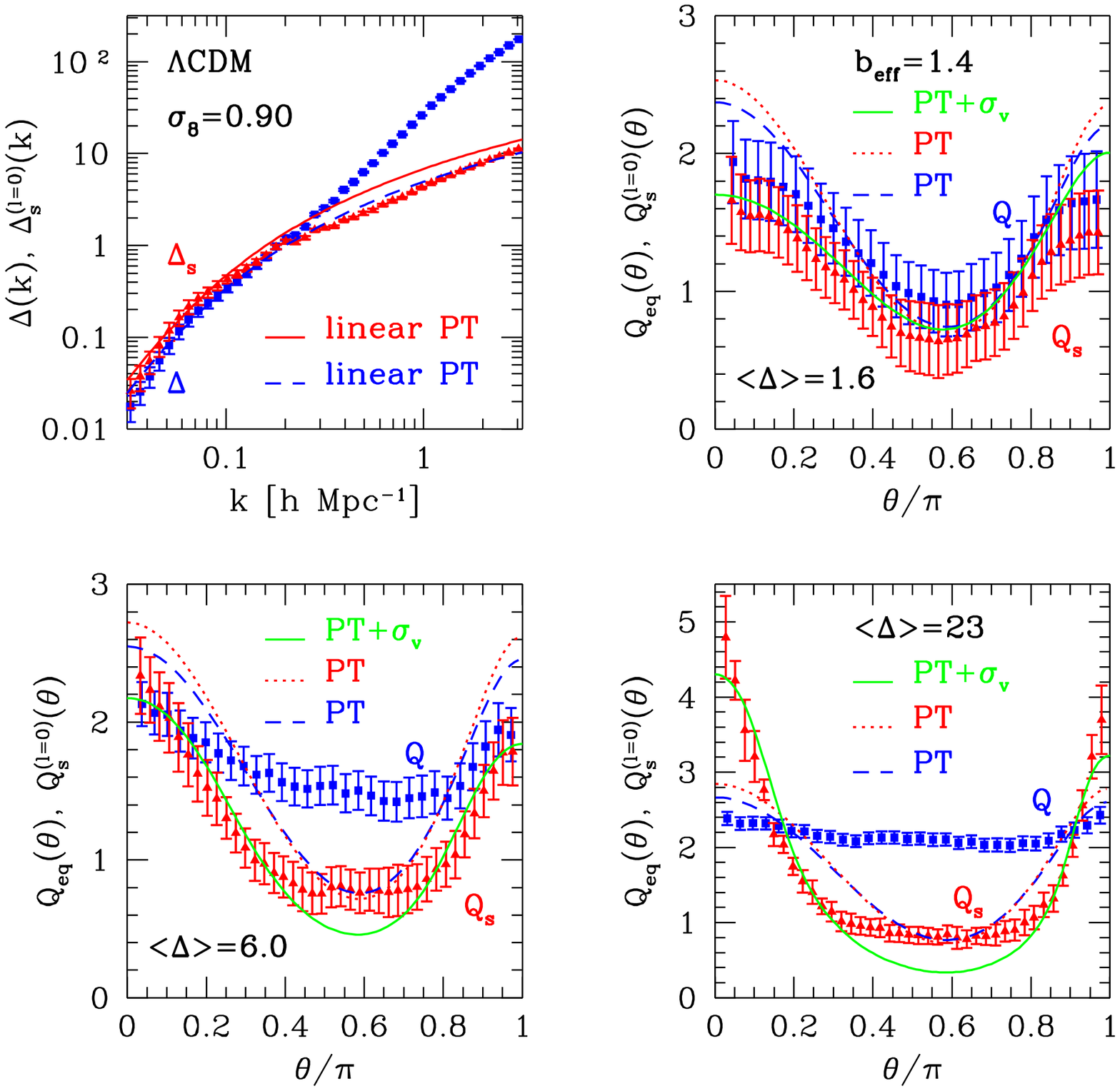,height=8.cm}}}
\caption{The top two panels show the quadrupole to monopole ratios
$R_{\rm P}$ and $R_{\rm B}$ in SCDM (top) and $\Lambda$CDM
simulations. The full lines show the predictions of PT convolved with
an exponential velocity dispersion model,
Eqs.~(\protect\ref{Ppheno}-\protect\ref{Bpheno}).
The bottom four panels show the power spectrum and the hierarchical
amplitude $Q$ in real (squares) and redshift (triangles) space for
$k_2/k_1$ configurations as a function of the angle $\theta$ between
$\k_1$ and $\k_2$ for three different scales.}
\end{figure}

\vfill
\end{document}